\documentstyle[twocolumn,aps,epsfig,floats]{revtex}
\draft
\begin{document}
\twocolumn[\hsize\textwidth\columnwidth\hsize\csname @twocolumnfalse\endcsname

\title{Bulk-sensitive photoemission spectroscopy of  
$\bf \rm A_{2}FeMoO_{6}$ double perovskites (A=Sr, Ba)}

\author {J. -S. Kang$^1$, J. H. Kim$^1$, 
	A. Sekiyama$^2$, S. Kasai$^2$, S. Suga$^2$, 
	S. W. Han$^3$, K. H. Kim$^3$, 
	T. Muro$^4$, Y. Saitoh$^5$, 
	C. Hwang$^6$, C. G. Olson$^7$, 
	B. J. Park$^8$, B. W. Lee$^8$, 
	J. H. Shim$^9$, J. H. Park$^9$, and B. I. Min$^9$}

\address{$^1$Department of Physics, The Catholic University of Korea,
        Puchon 420-743, Korea}

\address{$^2$Department of Material Physics, Graduate School 
	of Engineering Science, Osaka University,
        Osaka 560-8531, Japan}

\address{$^3$Department of Physics, Gyeongsang National University,
        Chinju 660-701, Korea}

\address{$^4$Department of Synchrotron Radiation Research, 
	Japan Atomic Energy Research Institute, SPring-8, 
        Hyogo 679-5198, Japan}

\address{$^5$Department of Synchrotron Radiation Research, 
	Kansai Research Establishment, 
	Japan Atomic Energy Research Institute, SPring-8, 
	Hyougo 679-5198, Japan}

\address{$^6$Korea Research Institute of Standards and Science,
        Taejon 305-600, Korea}

\address{$^7$Ames Laboratory, Iowa State University,
        Ames, Iowa 50011, U.S.A.}

\address{$^8$Department of Physics, Hankuk University of Foreign
	Studies, Yongin, Kyungki 449-791, Korea}

\address{$^9$Department of Physics, Pohang University of Science and
	Technology, Pohang 790-784, Korea}

\date{\today}
\maketitle

\begin{abstract}


Electronic structures of $\rm Sr_{2}FeMoO_{6}$ (SFMO) and
$\rm Ba_{2}FeMoO_{6}$ (BFMO) double perovskites have been
investigated using the Fe $2p \rightarrow 3d$ resonant photoemission 
spectroscopy (PES) and the Cooper minimum in the Mo $4d$ 
photoionization cross section.
The states close to the Fermi level 
are found to have strongly mixed Mo$-$Fe $t_{2g}$ character,
suggesting that the Fe valence is far from pure $3+$. 
The Fe $2p_{3/2}$ XAS spectra indicate 
the mixed-valent Fe$^{3+}-$Fe$^{2+}$ configurations, 
and the larger Fe$^{2+}$ component for BFMO than for SFMO,
suggesting a kind of double exchange interaction.
The valence-band PES spectra reveal good agreement with the LSDA+$U$ 
calculation.
\end{abstract} 

]

\narrowtext
   
Room-temperature magnetoresistance (MR) has been recently observed 
in the double-perovskite oxides of $\rm A_{2}BB^{\prime}MoO_{6}$ 
(A=Sr, Ba;B=Fe, B$^{\prime}$=Mo) with very high magnetic transition 
temperatures $\rm T_C$ ($\simeq 330-450$ K) 
\cite{Koba98,Borges99,Mori00a}.
A fundamental question in the double perovskites is the origin
of the high $\rm T_C$ ferrimagnetism and the low field MR.
Magnetization data for $\rm Sr_{2}FeMoO_{6}$ (SFMO) 
indicate ferrimagnetic coupling between
Fe$^{3+}$ and Mo$^{5+}$ ions \cite{Gala66},
and the MR was interpreted as due to intergrain tunneling with
the half-metallic electronic structure 
\cite{Koba98,Kim99,Tomioka00}. 
Neutron diffraction and M$\rm\ddot{o}$ssbauer spectroscopy on 
$\rm A_{2}FeMoO_{6}$ (AFMO) indicated an Fe moment of $4.0 
\sim 4.1 ~\mu_B$ but rather small localized moment on Mo, 
$-0.2 \sim -0.5~\mu_B$ \cite{Borges99,Garcia99,Mori00b}.
The ferrimagnetic coupling between Fe$^{3+}$ and Mo$^{5+}$ can be
understood in terms of the superexchange through the Fe-O-Mo 
$\pi$-bonding.
The superexchange model, however, is not compatible with the metallic 
nature of SFMO and $\rm Ba_{2}FeMoO_{6}$ (BFMO).
Further, the valence states of Fe and Mo ions are still controversial; 
some works are interpreted as the Fe$^{3+}$ \cite{Koba98,Ray01},
while other works favor the mixed valence of Fe$^{3+}-$Fe$^{2+}$ 
\cite{Garcia99,Linden00,Chmai00}.

Photoemission spectroscopy (PES) can provide direct information on 
the electronic structures of the double perovskites,
but no valence-band PES study on SFMO has been reported so far 
\cite{Kang01b,Sarma00}. 
In this paper, we report the first bulk-sensitive high-resolution 
valence-band PES study for SFMO and BFMO, including resonant 
photoemission spectroscopy (RPES) near the Fe $2p$ absorption edge 
and the O $1s$ and Fe $2p$ X-ray absorption spectroscopy (XAS) 
measurements. 


Polycrystalline SFMO and BFMO samples were prepared by the standard 
solid-state reaction method \cite{Kang01b}. 
The measured saturation 
magnetic moments are larger than $3.8 ~\mu_B$ (BFMO)
and $3.4 ~\mu_B$ (SFMO) per formula unit (f.u.), 
reflecting the well ordered Fe and Mo ions at B and B$^{\prime}$ sites. 
High-resolution Fe $2p \rightarrow 3d$ RPES experiments were 
performed at the BL25SU of SPring-8.
\cite{Saitoh00}.
Samples were fractured and measured in vacuum with a base pressure 
better than $\rm 3 \times 10^{-10}$ Torr at $\rm T \lesssim$ 20 K.
The Fermi level $\rm E_F$ of the system was determined from the valence 
band spectrum of a scraped Pd sample.
The total instrumental resolution [FWHM : full width at half maximum]
was about $130$ meV at $h\nu\sim 700$ eV.
All the spectra were normalized to the photon flux estimated
from the mirror current.  
Low energy PES experiments were carried out at the Ames/Montana 
beamline at the Synchrotron Radiation Center (SRC)\cite{Kang01b}.

\begin{figure}[hbt]
\epsfig{file=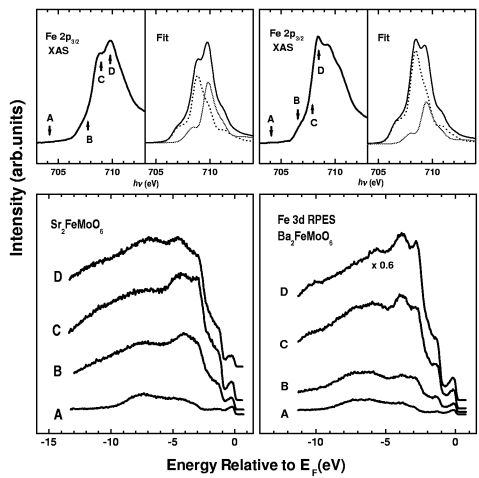,width=7.65cm}
\caption{Top: The measured Fe $2p_{3/2}$ XAS spectra (left) 
        and the fitting results (right) for SFMO and BFMO (see the text).
	Bottom: Valence-band spectra of SFMO and BFMO
	near the Fe $2p \rightarrow 3d$ absorption edge. }
\label{rpes}
\end{figure}

Figure~\ref{rpes} shows the RPES valence-band spectra of SFMO 
and BFMO near the Fe $2p_{3/2}$ absorption edge.
The top panels show the measured Fe $2p_{3/2}$ XAS spectra (left)
and the calculated XAS spectra (right).
The arrows in the XAS spectra represent $h\nu$'s where 
the spectra are obtained. 
For both SFMO and BFMO, the features near $\rm E_F$ 
($\sim -0.3$ eV), at $\sim -1.5$ eV, at $-3$ eV, and $-4$ eV 
in the valence-band spectra are enhanced greatly across the Fe $2p$ 
absorption edge, 
indicating that these states have large Fe $3d$ character.
As to the resonating behavior, the Fe $3d$-derived states at 
$-0.3$ eV and at $-1.5$ eV are due to 
the $t_{2g}^{x}\downarrow$ and $e_{g}^2\uparrow$ states, respectively
($0 < x \lesssim 1$) \cite{Kang01b}. 
These RPES measurements reveal that the high binding energy (BE) 
features ($-3\sim -8$ eV) also have the substantially large Fe $3d$ 
electron character, which is strongly hybridized with the O $2p$ 
electrons. 
In the above $h\nu$ region, however, the features for E $\lesssim -5$ 
eV are obscured by the overlapping broad Fe LMM Auger emission, 
which is evident as the non-flat slope for E $\lesssim -8$ eV 
for $h\nu\gtrsim 708$ eV. 

Our Fe $2p_{3/2}$ XAS spectrum for SFMO is basically similar to 
those of previous reports \cite{Ray01,Moreno01}. But the low-$h\nu$ 
shoulder around $h\nu \sim 707$ eV, which is present in our XAS 
spectra of both SFMO and BFMO, is missing in the existing data,
which is due to the better resolution ($\sim 100$ meV) 
employed in our XAS spectra.
It is evident that the relative intensities between the two 
peaks of the $2p_{3/2}$ absorption maxima ($h\nu\approx 708$ eV, 
710 eV) are reversed in SFMO and BFMO; 
namely a stronger peak is located at a higher $h\nu$ for SFMO 
but at a lower $h\nu$ for BFMO.
We believe that this difference is intrinsic since both samples 
showed a clean single peak in the O $1s$ core-level spectrum, 
and essentially the same XAS spectrum was reproduced 
for BFMO with different fractures. 
It has been known that the $2p_{3/2}$ absorption edge of Fe$^{2+}$ ions in the
octahedral symmetry exhibits a main peak at a lower $h\nu$
than that of Fe$^{3+}$ ions \cite{Laan92,Croc95}. 
Thus the difference in the Fe $2p_{3/2}$ XAS spectra between SFMO 
and BFMO suggests that Fe ions in BFMO have a larger $2+$ component 
than in SFMO.  
One possible reason for such a difference might be the different 
oxygen relaxation between Fe and Mo due to different
ion radii of Ba and Sr ions. Indeed the structural data provides 
that the relative Fe-O bond length is smaller in SFMO to have higher 
Fe valence state, than in BFMO \cite{Mori00b,Ritter00}.

In order to estimate the Fe valences of SFMO and BFMO in the ground 
states, we have analyzed the XAS spectra of SFMO and BFMO within 
the configuration interaction cluster model
where the effects of the multiplet interaction, the crystal field, 
and the hybridization with the O $p$ ligands are included 
\cite{deGroot,Min02}.
The calculated XAS spectra are shown in the top-right panels,
where dotted lines, grey lines, and solid lines denote the Fe$^{2+}$ 
and Fe$^{3+}$ component, and their sum, respectively.
This analysis shows that the ground states of both SFMO and BFMO 
have the $\rm Fe^{3+}-Fe^{2+}$ mixed states, but that BFMO has 
the larger Fe$^{2+}$ component.
If we use the Fe$^{3+}$ component only ($d^5$ and 
$d^6\underbar{L}^{1}$, $\underbar{L}^{}$: a ligand hole),
the experimental features of the reversed XAS lineshape for BFMO 
and the low-$h\nu$ shoulders ($h\nu \sim 707$ eV)
cannot be reproduced. 
This analysis implies that the ground states of SFMO and BFMO
are strongly mixed-valent with the $\rm Fe^{3+}-Mo^{5+}$ 
and $\rm Fe^{2+}-Mo^{6+}$ configurations. 

\begin{figure}[hbt]
\epsfig{file=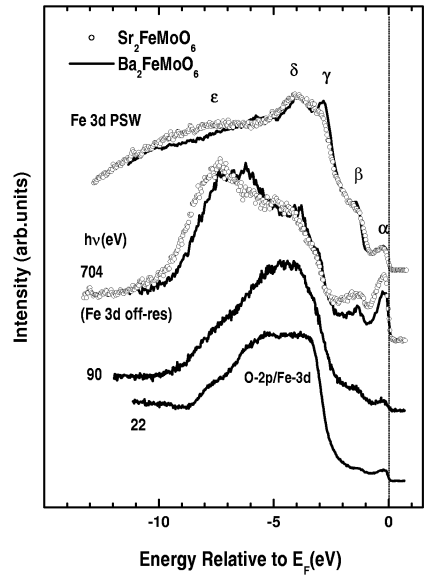,width=7.65cm}
\caption{Comparison of the valence-band spectra of SFMO (dots) 
	and BFMO (solid lines) for a wide $h\nu$ range.} 
\label{edc}
\end{figure}

Figure~\ref{edc} compares the valence-band spectra of SFMO  
and BFMO for a wide $h\nu$ range (22 eV $\le h\nu 
\lesssim 700$ eV).
The top two spectra are the differences between the on-resonance 
($h\nu \approx 708$ eV)
and off-resonance spectra in the Fe $2p \rightarrow 3d$ RPES, 
which can be considered as representing the Fe $3d$ 
partial spectral weight (PSW) distributions.
If one assumes Fe$^{3+}$ ($3d^5$), Mo$^{5+}$ ($4d^1$), and the filled
O $2p$ bands ($2p^6$) in AFMO, then the cross-section ratio of 
Fe $3d$ : Mo $4d$ : O $2p$ per unit cell is about 
$\sim 4 \%$ : $\sim 5 \%$ : $\sim 91 \%$ at $h\nu\approx 20$ eV, 
$\sim 25 \%$ : $\sim 0 \%$ : $\sim 75 \%$ at $h\nu\approx 90$ eV,
and $\sim 41 \%$ : $\sim 9 \%$ : $\sim 49 \%$ at $h\nu\approx 700$ eV
according to an atomic photoionization cross-section calculation
\cite{Yeh85}.
The $h\nu=22$ eV spectrum can be considered as the O $2p$ PSW
because of the dominant O $2p$ emission at $h\nu=22$ eV. 
With increasing $h\nu$, the Fe $3d$ and Mo $4d$ emissions increase
with respect to the O $2p$ emission, 
except for the $h\nu$ region corresponding to the Cooper minimum 
in the Mo $4d$ cross section around $h\nu\sim 90$ eV, 
and the Fe $3d$ and O $2p$ emissions become comparable at 
$h\nu\approx 700$ eV.
Consequently, the Mo $4d$ emission is negligible at $h\nu\sim 90$ eV,
and becomes the largest in the Fe $3d$ off-resonance spectrum 
($h\nu=704$ eV).
Thus the pronounced feature at $\rm E_F$ at $h\nu=704$ eV is due to 
the larger Mo $4d$ contribution than at other $h\nu^{\prime}$s, 
and may also represent the intrinsic feature in the bulk PES spectra 
\cite{Seki00}.

The valence-band spectra of SFMO and BFMO share very common features. 
First, metallic Fermi edges are observed in both systems at T $\sim 20$ K, 
which confirms the metallic behavior at low T.
Secondly, several structures appear at similar energies, labeled as 
$\alpha$ ($-0.3$ eV), $\beta$ ($-1.5$ eV), $\gamma$ ($-3.5$ eV),
$\delta$ ($-4$ eV), and $\epsilon$ ($\sim -8$ eV).
Based on the finding in Fig.~\ref{rpes},
the features of $\alpha, ~\beta, ~\gamma, ~\delta$ are identified
to have large Fe $3d$-derived electron character.
Note that the peak $\alpha$ is much larger at $h\nu=704$ eV than 
at $h\nu=90$ eV (Cooper minimum), indicating that the peak 
$\alpha$ has predominant Mo $4d$ electron character.
Therefore this figure provides experimental evidence that
the peak $\alpha$ has both Mo $4d$ and Fe $3d$ electron character,
but with larger Mo $4d$ character.  

\begin{figure}[hbt]
\epsfig{file=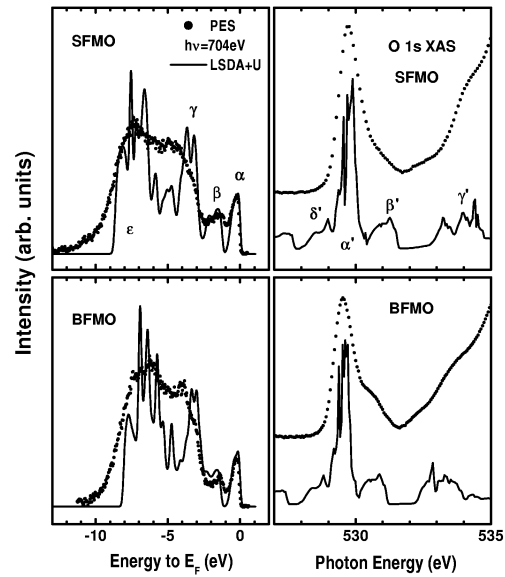,width=7.65cm}
\caption{Left: Comparison of the $h\nu=704$ eV valence-band spectra 
        (dots) to the weighted sum of the PLDOSs 
	(solid lines) obtained from the LSDA+$U$ calculation.
        Right: Comparison of the O $1s$ XAS spectra to the sum of 
	the Fe $3d$, Mo $4d$, O $2p$, and Sr(Ba) $s/p$ PLDOSs.}
\label{comp1}
\end{figure}

We now discuss the differences in the valence-band spectra
between SFMO and BFMO. 
First, the $\alpha$ peak at $\rm E_F$ for BFMO is weaker
than that in SFMO with respect to the $\epsilon$ peak,
and secondly the band width of the O $2p-$Fe $3d$ peak in BFMO
is narrower than that in SFMO. 
The former difference suggests that the Mo $4d$ electron occupancy
in BFMO is lower than that in SFMO since $\alpha$ has both Fe $3d$
and Mo $4d$ electron character but the intensity of the peak $\alpha$
in the Fe $3d$ PSW is the same for SFMO and BFMO. 
Accordingly, this finding indicates 
larger $\rm Fe^{2+}-Mo^{6+}$ character in BFMO than in SFMO, 
which is consistent with that of
the Fe $2p_{3/2}$ XAS spectra (Fig.~\ref{rpes}).
The difference in the bandwidth is in agreement with the LSDA+$U$
($U$: the Coulomb correlation parameter) 
calculations shown in Fig.~\ref{comp1}, and reflects the larger 
Fe-O-Mo bond length in BFMO due to its larger lattice constant 
\cite{Ritter00}.

The left panel of Fig.~\ref{comp1} compares the $h\nu=704$ eV 
valence-band spectra (dots) to the weighted sum of the PLDOSs 
(projected local densities of states) of Fe $3d$, Mo $4d$, and O $2p$ 
states (solid lines). 
The PLDOSs have been obtained from the LSDA+$U$ calculation 
\cite{Kang01b}. 
The parameters used in this calculation are $U=3.0$ eV and $J=0.97$ eV 
($J$: the exchange correlation parameter) for Fe $3d$ electrons. 
For the comparison to the valence-band spectra, 
the summed PLDOSs below $\rm E_F$ have been convoluted with a Gaussian 
function of 0.2 eV at FWHM to simulate the instrumental resolution. 
The right panel compares the O $1s$ XAS spectra to the sum of 
the unoccupied part of the Fe $3d$, Mo $4d$, O $2p$, and Sr (Ba) 
$s/p$ PLDOSs. The O $1s$ XAS spectrum provides a reasonable 
approximation for representing the unoccupied conduction bands 
since it reflects the transition from the O $1s$ to the unoccupied 
O $2p$ states hybridized to the other electronic states.
The calculated integer total magnetic moments 
($4.0 ~\mu_{B}$ per f.u.) 
are consistent with the half-metallic nature of SFMO and BFMO. 

In the valence-band spectra, the lowest energy peak of $\alpha$ 
arises from the Mo $t_{2g}\downarrow-$ Fe $t_{2g}\downarrow$ 
hybridized states 
with the occupied bandwidth of $\sim 1$ eV.
The features of $\beta$ and $\gamma$ have mainly
Fe $e_{g}\uparrow$ and Fe $t_{2g}\uparrow$ character, respectively. 
The high BE feature $\epsilon$ originates from 
the Mo $4d-$ Fe $3d-$ O $p$ hybridized bonding states.
In the XAS spectra, the low energy peak $\alpha^{\prime}$ 
has mainly Mo $t_{2g}\uparrow$ and Fe $t_{2g}\downarrow$ character. 
The feature of $\beta^{\prime}$ originates from
Fe $e_{g}\downarrow$ states, 
and $\gamma^{\prime}$ from Mo $e_{g}\uparrow \downarrow$ states.
The weak feature $\delta^{\prime}$ is due to Mo $t_{2g}\downarrow$ 
states. 
It is clearly shown that the LSDA+$U$ calculations agree
quite well with the valence-band spectra, in the peak positions
and the band widths, and also explain the low energy features of 
the XAS spectra reasonably well. 
If we shift the O $1s$ XAS spectra by referring to the LSDA+$U$
calculations \cite{Kang01b,shift}, 
the energy separations between the lowest energy peaks in the 
valence-band PES ($\alpha$) and those in the O $1s$ XAS 
($\alpha^{\prime}$) spectra become about $\sim 1$ eV. 
This value is comparable to, but slighter larger than the low energy 
peak observed in the optical conductivity spectrum for SFMO 
\cite{Tomioka00}, which was ascribed to the Fe $e_{2g}\uparrow 
~\rightarrow$ Mo $t_{2g}\uparrow$ interband transition. 
We, however, ascribe this energy separation 
to the energy difference between the occupied Mo $t_{2g}\downarrow-$Fe 
$t_{2g}\downarrow$ hybridized state 
and the unoccupied Fe $t_{2g} \downarrow$ state mixed with
O $p-$Mo $t_{2g}$ states. 

We have found that the Fe $t_{2g}\downarrow$ and Mo $t_{2g}\downarrow$ 
bands are almost degenerate and so $t_{2g}$ electrons become itinerant,
suggesting that two degenerate valence 
states $\rm Fe^{3+}-Mo^{5+}$ and $\rm Fe^{2+}-Mo^{6+}$ would produce 
a type of double exchange (DE) interaction \cite{Sleight72}. 
That is, hopping of itinerant $t_{2g}\downarrow$ electrons 
between Fe and Mo sites yields a kinetic energy gain to 
induce ferrimagnetism between Fe and Mo spins
and the half metallicity in SFMO and BFMO.  Therefore Fe 
and Mo ions in SFMO and BFMO do not have definite valence states
\cite{Linden00,Chmai00}.
The difference from the case of CMR manganites is that the spins of 
itinerant carriers are opposite to the localized 
Fe spins ($t_{2g}^3\uparrow e_{g}^2\uparrow$).
Sarma {\it et al.} \cite{Sarma00} and Fang {\it et al.} \cite{Fang01}
have recently proposed alternative models
to explain the magnetism in double perovskites. 
They regard the double perovskite system as a ferromagnet not a ferrimagnet.
That is, originally nonmagnetic
Mo $t_{2g}$ electrons are antiferromagnetically 
polarized by the ferromagnetically ordered 
$\rm Fe^{3+}$ spins through the hybridization between Fe $3d$ and Mo $4d$
states, thereby the resulting kinetic energy gain stabilizes the
ferromagnetic state. 
In view of the role of kinetic energy optimization,
this mechanism seems similar to the DE mechanism. 
However, using their models, it is hard to explain the ferrimagnetism
observed in the insulating double perovskites 
such as $\rm Ca_2FeReO_6$ and Ba$_2B$ReO$_6$ ($B$=Mn, Co, Ni) 
which have one more $d$ electron in the B$^{\prime}$ site \cite{Sleight72}.

Ferrimagnetism in the insulating double perovskites of
$\rm A_{2}BB^{\prime}O_{6}$
is better described in terms of the superexchange interaction
through B-B$^{\prime}$-O $\pi$ orbitals.
In the metallic SFMO and BFMO systems,
the DE-like interaction becomes operative due to
the degenerate Fe and Mo $t_{2g} \downarrow$ states near $\rm E_F$.
In the DE mechanism, the magnetic transition temperature $\rm T_C$
is proportional to the hopping strength, i.e., the bandwidth 
of the itinerant $t_{2g}\downarrow$ states \cite{Chatt00}.
The observed correlation between $\rm T_C$ and the estimated 
bandwidth in metallic double perovskites
provides evidence for the operation of the DE-like
interaction \cite{Ritter00}.
Noteworthy is that both the superexchange and the DE interactions
induce the ferrimagnetism in double perovskites, differently from
the case of CMR manganites in which two interactions are competitive.
Further, the Jahn-Teller effect would be weaker in double perovskites 
due to relevant $t_{2g}$ orbitals 
near $\rm E_F$, as compared to $e_{g}$ orbitals in manganites.
Hence $\rm T_C$ can be high for SFMO and BFMO. 


In conclusion,
using the bulk-sensitive Fe $2p \rightarrow 3d$ RPES and O $1s$ XAS,
the states close to $\rm E_F$ are found to be predominantly 
of Mo $t_{2g}\downarrow$ and Fe $t_{2g}\downarrow$ character,
indicating that the Fe valence is far from pure $3+$ ($d^5$). 
The Fe $2p_{3/2}$ XAS spectra shows that the ground states 
of both SFMO and BFMO consist of the strongly-mixed 
Fe$^{3+}-$Fe$^{2+}$ configurations, and that BFMO has 
the larger Fe$^{2+}$ component.
The measured valence-band spectra agree very well with the LSDA+$U$ 
calculation, suggesting that the DE-like interaction is operative 
to produce the half metallicity and ferrimagnetism in SFMO
and BFMO.

Acknowledgments$-$
Helpful discussions with D.D. Sarma are greatly appreciated.
This work was supported by the KOSEF through the CSCMR
at SNU and the eSSC at POSTECH.
PES experiments were performed at the SPring-8
(JASRI: 2000B0028-NS-np) and at the SRC (NSF: DMR-0084402).

\begin{references}
\bibitem{shift} The O $1s$ XAS spectra of SFMO and BFMO are shifted
	by $-528.7$ eV and $-528.5$ eV, respectively. 
	There is uncertainty in choosing the zero positions in the 
	XAS spectra.
\end{references}
\end{document}